\documentclass[english,aps,pra,reprint,noshowpacs,superscriptaddress]{revtex4-1}   
\usepackage[T1]{fontenc}	
\usepackage[latin9]{inputenc}	
\usepackage{geometry}		
\usepackage{graphicx}
\usepackage{xcolor}
\usepackage[above,below]{placeins}	
\usepackage{times}
\usepackage{enumitem}
\usepackage{hyperref}  
\hypersetup{colorlinks=true,urlcolor=blue,citecolor=blue,linkcolor=blue}   
\begin{document}

\title{Pathways to proposing causes for unexpected experimental results}
\author{Laura R\'{i}os}
\author{Benjamin Pollard}
\author{Dimitri R. Dounas-Frazer}
\author{H.~J. Lewandowski}
\affiliation{Department of Physics, University of Colorado Boulder, Boulder, CO, 80309}
\affiliation{JILA, National Institute of Standards and Technology and University of Colorado Boulder, Boulder, CO, 80309}

\begin{abstract}
	Models of physical systems are used to explain and predict experimental results and observations. When students encounter discrepancies between the actual and expected behavior of a system, they revise their models to include the newly acquired observations, or change their apparatus to better represent their models. The Modeling Framework for Experimental Physics describes the process of matching measurements and observations to models by making revisions to resolve discrepancies. As part of a larger effort to create assessments of students' modeling abilities in the context of upper-division electronics courses, we used the Modeling Framework to develop and code think-aloud problem-solving activities centered on troubleshooting an inverting amplifier circuit. We observed that some participants iteratively and continuously made measurements and revisions if they could not immediately propose a cause for an observed discrepancy. This pathway has not been previously discussed in the Modeling Framework. In this paper, we discuss two episodes where students undergo this process to converge on a proposed cause \textit{post hoc}. We conclude by discussing implications for a modeling assessment based on the observed modeling behavior.
\end{abstract}

\maketitle

\section{Introduction}
\vspace{-1.0em}

According to the recommendations released by the American Association of Physics Teachers (AAPT), modeling--the construction, testing, use, and revision of models of physical phenomena and apparatus--should be a focus of physics laboratory courses \cite{AAPT2015}. There has been a considerable amount of research at both the introductory \cite{etkina,Megowan-Romanowicz2011} and upper-division level \cite{Zwickl2015, Stanley2017a} on how students engage in model-based reasoning. The Modeling Framework for Experimental Physics (Fig. \ref{fig:framework}), developed by our group, describes the process by which physicists bring measurements and predictions from models into agreement \cite{Zwickl2015}. The Framework is composed of several subtasks, depicted in gray boxes in Fig. \ref{fig:framework}: making measurements, constructing models, making comparisons between data and predictions to assess discrepancies, proposing causes for those discrepancies, and enacting revisions to resolve them. The Framework has been used as both a research tool to understand students' engagement in modeling and as a guide to introduce modeling into physics lab courses \cite{Zwickl2014,Dounas-Frazer2016a}. 

Although modeling is a core scientific practice in experimental physics, and a major learning outcome for lab courses, no validated instruments exist for assessing model-based reasoning in labs at the upper-division level. Additionally, the National Research Council (NRC) has recently called for increased attention to assessments of experimental physics practices \cite{NRC2013}. In response to the AAPT recommendations and the NRC's call to action, our group is currently in the process of developing validated and scalable assessments of students' experimental modeling abilities. The assessment development has four phases: 1) provide domain-specific test objectives \cite{Dounas-Frazer2017d}; 2) characterize how students navigate and justify their choices during think-aloud problem-solving (TAPS) interviews; 3) create a free-response assessment with input from expert physicists; and finally, 4) produce a coupled-multiple response assessment. 

Phase 1 of the process yielded test objectives and relevant experimental contexts from interviews with electronics and optics lab instructors across various institutions \cite{Dounas-Frazer2017d}. Using this information, we developed a TAPS interview protocol for Phase 2. One of the goals of this phase was to identify the order in which students engaged in the modeling subtasks (Fig. \ref{fig:framework}) when working with an electric circuit. Here, we report on a subset of results from Phase 2, where we focus on students' proficiency with identifying possible causes of a discrepancy between a measurement and a prediction from a model, and when in the process they articulate the proposed causes. 

This work was motivated, in part, by previous studies into students' modeling abilities. In one study, Zwickl \textit{et al.} \cite{Zwickl2014} found that students often do not propose a cause when confronted with unexpected results. This theme was echoed by electronics and optics physics instructors \cite{Dounas-Frazer2017d}. Another study on how students document modeling in their lab notebooks found that even when the modeling process is scaffolded by course materials, the students ``generally did not provide [in their lab notebooks] actionable ways of implementing these proposed revisions'' \cite{Stanley2017a}. Thus, there is demonstrated need to understand the circumstances under which a student does or does not propose causes to motivate revisions, as it is a critical subtask during the modeling process. 

In the work presented here, we describe our progress towards characterizing the pathways students take during the modeling process, in particular around the Propose Cause subtask. Specifically, we expound upon two episodes in which students proposed a cause for a discrepancy after continuously and recursively revising their apparatus, making measurements, and comparing the results of their revisions. This newly observed pathway has implications for the assessment design and adds to the understanding of students' engagement with a core scientific practice.

\section{Methods and Participants}
\vspace{-1.0em}
We conducted 10 TAPS interviews with upper-division physics and engineering physics undergraduate students at the University of Colorado Boulder (CU Boulder). We designed an activity with the aim of examining student reasoning around the operational limits of op-amps, while allowing for several model and apparatus revisions. The main focus of the activity was a circuit composed of an LF356 op-amp chip in an inverting amplifier circuit with a gain of 10. At the onset of the activity, the circuit was already powered, with the frequency ($f$), input voltage ($V_{\text{in}})$, and power rails ($V_{+/-}$) preset to $V_{\text{in}} = 3 V_{\text{pp}}$, $f = 250$ kHz, and $V_{{+/-}} = \pm 10$ V. We designed the initial conditions such that the output waveform would be significantly clipped, and would appear slightly distorted due to slew rate limit of the op-amp, but not enough that the signal would be unrecognizable. The slew rate limit of the op-amp determines the maximum rate of change of the output voltage, and depends on both the frequency and output voltage amplitude \cite{Horowitz:1989:AE:76734}. Operating above the slew rate limit will result in an output voltage waveform that appears closer to a sawtooth than a sine wave, and with a smaller amplitude. 

At the start of the activity, we provided the student with a schematic diagram of the circuit, a data sheet for the op-amp, and a prebuilt functioning circuit. They were also given the equation for the gain, \textit{A}, of the circuit as a function of the resistance of the feedback and input resistors, i.e. $A = R_f/R_{in}$. The interviewer read a short prompt to the students before they began their work on the activity. The goal of the activity was framed as follows:

\begin{itemize}
 \item[] \textit{Your goal is to match the output waveform to the input waveform with the correct gain, phase, and shape. By the `correct gain,' I mean that the output waveform should be amplified by the gain you predict from the ratio of the resistors. You can change the value of A [the gain] if you deem it necessary to achieve your goal.} 
\end{itemize}

The activity ended when the student declared that the gain, shape, and phase were correct, or until they had worked on the circuit for about 45 minutes. After the student completed the activity, they were asked several follow-up questions, which were used to clarify revisions and measurements they made during the activity, and to understand the student's background working with electronics. The TAPS interviews were audio and video recorded. Each interview lasted between 55 - 70 minutes.

We recruited junior and senior students who had taken the electronics lab (PHYS 3330: Electronics for Physical Sciences) in the CU Physics department in Fall 2016, Spring 2017, or Fall 2017. The interview took place during Fall 2017. Students were compensated monetarily for their time. Ten students agreed to participate; five were juniors and five were seniors at the time of the activity. Eight of the 10 agreed to answer our final demographic question (``Finally, is it all right if you tell me your gender and your race and/or ethnicity?''). Of those, 8 identified as white, 6 identified as male, and 2 identified as female. We do not report intersections of demographic information to protect the identity of our student participants. 

\begin{figure}
	\begin{center}
		\includegraphics[width=0.8\linewidth]{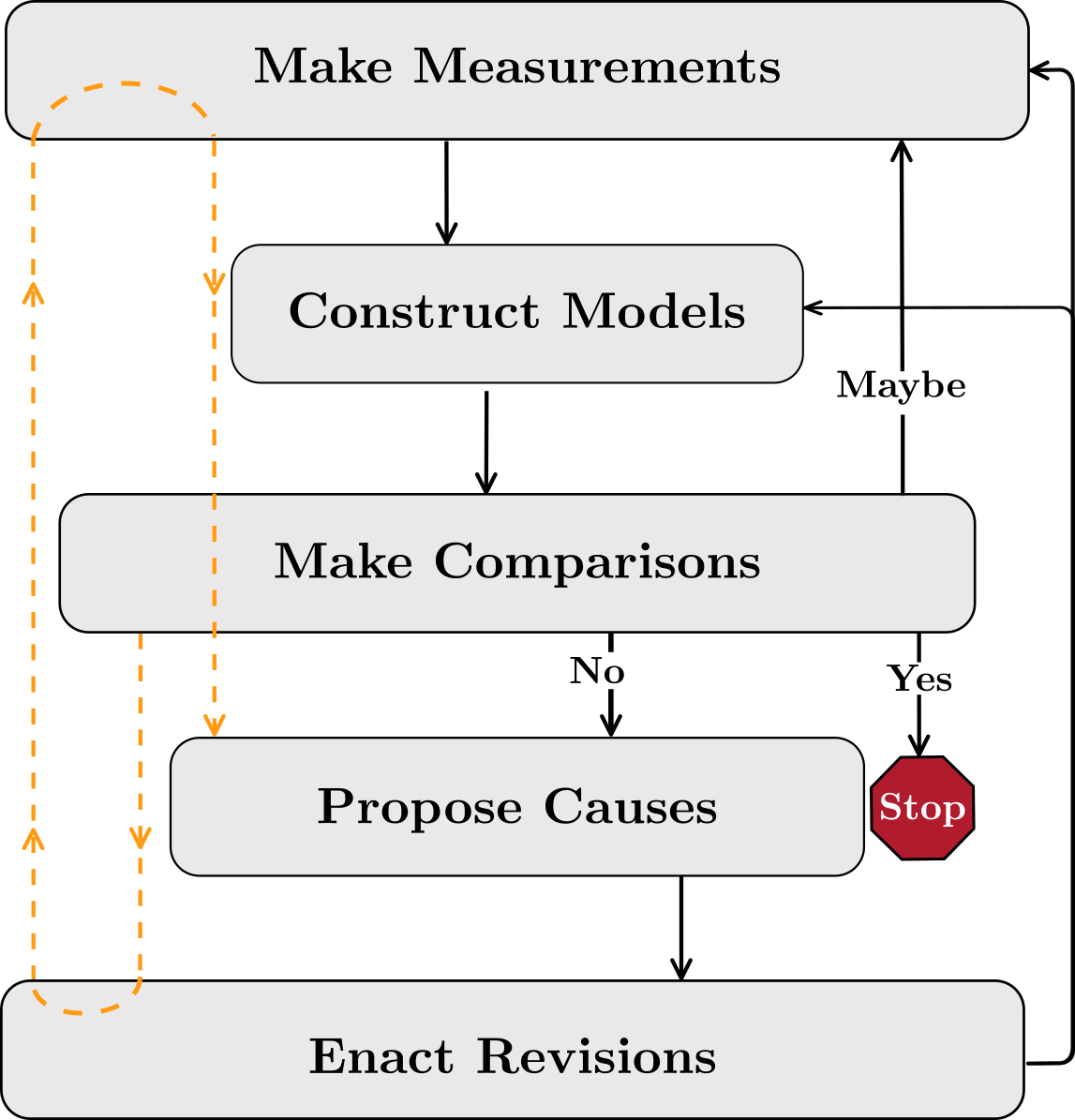}
		\caption{Modeling Framework for Experimental Physics: A condensed schematic of the Modeling Framework, originally conceptualized by Zwickl \textit{et al.} \cite{Zwickl2014} The subtasks were used as top-level codes. Specific outcomes and articulations were operationalized as subcodes. The orange dashed line represents the pathway we describe in this paper as an empirical extension of the Framework.}
		\label{fig:framework} 
	\end{center}
\end{figure}

Using the Modeling Framework as the primary basis for an \textit{a priori} coding scheme, we used the broad subtasks (gray boxes in Fig. \ref{fig:framework}) as the top-level codes, with subcodes for Enact Revisions, Propose Causes, Make Measurements, and Make Comparisons, which were defined based on the activity design, LR's experience in electronics, and input from all authors. For example, under ``Make Comparisons,'' the act of comparing the expected phase of an inverting amplifier circuit to the observed measurement was a subcode.

After the full coding scheme was created, we completed a training phase before the final coding. Authors LR and BP collaboratively coded a 10 minute segment of an interview. After discussing the coding scheme and making minor revisions to ameliorate possible vagueness, LR and BP then individually coded a separate 10 minute segment of a different interview. BP and LR reconciled differences, and discussed ways in which the robustness of the codebook could be improved, which resulted in minor changes to the code definitions. We completed the training and separate coding process for two iterations, afterwards getting a Cohen's kappa of 0.69, indicating substantial agreement. LR then coded all 10 interviews with the improved codebook.

\section{Results and Discussion}
\vspace{-1.0em}

The larger research study from which these data were extracted sought to understand 1) how long students spent in each subtask in the Modeling Framework, 2) in which order the students engaged in subtasks, and 3) what types of revisions, measurements, comparisons, and causes the students used. For this study, our analysis was driven by the second research question, exploring the order in which students perform modeling subtasks. Preliminary findings revealed that the students are often quick to identify discrepancies, but did not propose any probable causes to motivate a revision, opting instead to make iterative revisions and measurements with intermittent comparisons, eventually describing a cause for the discrepancy \textit{post hoc}.  Often, the \textit{post hoc} proposed cause is not the exact reason for the discrepancy, which highlights how difficult it is to propose causes. 

In the following examples, we expatiate two episodes from TAPS interviews with Lorca and Plath \footnote{All names are pseudonyms.}. They identify a discrepancy, but do not propose a probable cause initially. Instead, they engage in modeling by making measurements and constantly reflecting on the effect of their revisions. By measuring and revising continuously, they start to converge on possible causes for the discrepancy. These vignettes provide clear examples of this newly observed pathway. 

In one episode, student Plath sees an unexpectedly small gain, and reasons about circuit behavior at high frequencies. Before the quote below, Plath has identified that the gain should be 10 based on the values of the resistors, i.e. $A = R_f/R_{in} = 10$. From the ratio of $V_{\text{out}}$ to $V_{\text{in}}$ on the oscilloscope, she observes that the gain does not match her expectation, as it appears to be smaller than 10. At this point in the interview, the frequency is set at 1 MHz and the input voltage is $2 \; V_{\text{pp}}$. She checks a few aspects of the output signal, and correctly concludes that the circuit itself must be functional, but does not know why the gain appears to be smaller than it should be. She does not propose a cause based on her model of circuit behavior, so she starts out by revising the frequency on the function generator:

\begin{itemize}[noitemsep]
	\item[1] \textit{The signal still has a weird shape. [\ldots]
	\item[2] So definitely at 1 MHz, my gain decreased a lot, but is there also selectivity for low frequency? 
	\item[3] At 1 kHz, looks like my output is 2, 4, 6, 8, 10 [volts] and the input is about 1 [volt].  
	\item[4] Oh! So, the gain looks better now, at this frequency. So [\ldots] the output of the circuit is like right around 10. And the input is right around 1 volt peak-to-peak.  
	\item[5] So, what happens if I go to really low frequency? Like, 100 Hz? 
	\item[6] [pause] The gain is, again about 10 and the output is like 10 [volts] and the input is like 1 [volt], so, that's still good.
	\item[7] Maybe it acts like a low pass filter-ish behavior?}
	\item[] \textbf{Plath, 23:30--26:00}
\end{itemize}

First, Plath identifies a qualitative discrepancy in line 1 in the above quote. From the video data, it is evident that the ``weird shape'' she is referring to is the unexpected amplitude of the output signal, not the sinusoidal shape, or the phase. She correctly identifies that the gain should not be less than 10, but does not propose a cause initially. Instead, Plath comments on the initial conditions and subsequently decides to investigate directly the relationship between frequency and gain in line 2. To do so, she iteratively revises her measurement/test equipment (i.e. decreases the frequency) while simultaneously measuring (i.e. watching the ratio of $V_{\text{out}}$ to $V_{\text{in}}$ on the oscilloscope). Specifically, upon decreasing the frequency, Plath stops and reflects on the effect of her revision in lines 3-4, noticing that a lower frequency fixed the apparent small gain. Once Plath establishes the relationship between frequency and gain, she continues to revise the frequency setting to understand the boundaries of the frequency dependence in lines 5-6. 

Finally, she proposes a cause for the unexpectedly low gain \textit{post hoc} when she ends by interpreting the relationship between frequency and gain as a low-pass filter in line 12. A low-pass filter passes signals with frequencies lower than its cutoff frequency, $f_C$. Signals with $f > f_C$ will be attenuated \cite{Horowitz:1989:AE:76734}, so Plath correctly converged on a phenomenological proposed cause by generating specific knowledge about how frequency affects gain.  

In a distinct episode, Lorca has been systematically working through making sure the gain, phase, and shape are what he expects. After changing the gain to 1 to alleviate the clipping, he determines the gain and phase are correct, but is concerned that the shape is not good enough. Specifically, he observes unwanted phase-unsynced, high-frequency noise that appears as quick signals that travel through an otherwise stable $V_{\text{out}}$ signal. Lorca understands that the noise is undesirable, and like Plath, does not propose a cause immediately. Instead, he measures and revises until he sees that the waveform appears stabilized at lower frequencies. 

\begin{itemize}[noitemsep]
\item [] \textit{
	\item [1] And I didn't even think of trying to deal with all of this by lowering the frequency, which I'm doing now. 
	\item[2] And I'm doing this to try to get some of the pulses [noise] out of the waves. 
	\item[3] So I moved it from down to 50 kHz. [pause] And that really didn't help, but I thought that might help, just because...
	\item[4] Actually, it kind of helped a little bit. [\ldots] 
	\item[5] That's just because of the BNC cables... When you pass frequency through them, or, when you pass high frequencies through them, they turn into resistors, and I know that can mess with your input and output waves. So, keeping this [frequency] lower is probably more helpful. 
	\item[6] Yeah, I moved it down to 10 kHz...I think that actually seems to give a better output wave, just in terms of looking at the shape of the wave.}
	\item[] \textbf{Lorca, 37:30--40:00}	
\end{itemize}

 Lorca identifies the discrepancy by articulating that a functioning inverting amplifier should not have noise in lines 1-2. Then, he iteratively changes the frequency and reflects on how the decreased frequency appears to help the ``pulses'' through his revision, as seen in lines 1, 3-4, and 6. Like Plath, Lorca is measuring and revising to directly investigate how frequency affects the stability of the output signal. Then, Lorca reasons why decreasing the frequency alleviated the appearance of the ``pulses'' by reasoning through the effect of the characteristic impedance of the BNC cables in line 5. 

Here, we may infer that Lorca is describing what happens when one takes into account the characteristic impedance of the BNC cables connecting the circuit board to the function generator and oscilloscope, and attempting to rationalize the noise pick-up with a revised model of the BNC cables. Specifically, BNC cables have a small capacitance that becomes a significant impedance when the frequency is large. Lorca recognized that the BNC cables do not always behave as an ideal circuit element by invoking their characteristic impedance. However, the circuit being investigated here operates in a regime in which the capacitance of the BNC cables is negligible. So, the proposed cause Lorca converges upon does not explain the noise pick-up he observed. Compared to Plath's episode, Lorca's episode is more typical of the complete data set. Often, students are able to incorporate their observations into revised models in ways that are consistent with their previous knowledge, but nevertheless they do not converge upon the cause that describes the initial observed discrepancy. This trend spotlights that probable causes are difficult to propose even when drawing upon correct concepts.

\section{Implications for Assessment}
 \vspace{-1.0em}

The two episodes described here have concrete implications for an overall assessment of model-based reasoning in physics labs. Considering the overall context of the second phase of the modeling assessment development, which is generally concerned with which processes students take to engage with modeling, we conclude that our assessment must allow students to make measurements and revisions to reconcile unexpected experimental results. In particular, the continuous measurement process appears crucial for proposing causes \textit{post hoc} for observed discrepancies. Since students generally experience difficulties around the Propose Cause subtask, the assessments needs to allow students to skip the propose cause prompt, and then circle back around after more measurements and revisions. The assessment should not focus solely on whether a probable cause was proposed initially. The process by which a student arrives at a proposed cause should also be accommodated in the ultimate assessment.

More generally, these data illustrate the importance of focusing on process-based outcomes of the assessment, instead of task outcomes \cite{Dounas-Frazer2017d}. Future reports will detail other guiding principles for subsequent phases of the modeling assessment. 

\section{Conclusion}
\vspace{-1.0em}

We conducted 10 think-aloud problem-solving interviews with CU Boulder upper-division physics students in order to learn what processes they undertake while troubleshooting an inverting amplifier circuit operating outside its recommended limits. In this work, we show two examples of students who do not initially identify a probable cause for the discrepancies between experimental results and expectations. They proceed to complete constant and iterative measurements and revisions of their measurement/test apparatus to arrive at a proposed cause \textit{post hoc}. We argue that these data have strong implications for the modeling assessment we are currently developing. Specifically, the modeling assessment should allow a student to continue measuring and revising if they do not propose a cause for observed discrepancies initially, and allow them to propose a cause for the discrepancy \textit{post hoc}.   

\acknowledgments{\vspace{-1.0em}We are grateful to the CU Boulder physics students who participated in this study. We would also like to thank Dr. Kevin L. Van De Bogart for his suggestions on coding the data, and Charles Ramey, Drs. Gina M. Quan, Janet Y. Tsai, Jacob T. Stanley, and Joel C. Corbo for their feedback on the manuscript. This work is supported by the National Science Foundation under Grant Nos. DUE-1611868, DUE-1726045 and PHYS-1734006.}

\bibliographystyle{apsrev}  	
\bibliography{PERC2018}  	

\end{document}